\documentclass[article,12pt,a4paper,cyremdash,oneside]{ncc}

\usepackage{cmap}
\usepackage{concmath}
\usepackage{amsmath,amssymb}
\usepackage{cite,fancyvrb,icomma,indentfirst}
\usepackage[utf8]{inputenc}
\usepackage[T1,T2A]{fontenc}
\usepackage[english,russian]{babel}

\FromMargins[h]{3cm}{2cm}{1cm}{2cm}
\fvset{fontsize=\small,frame=leftline,framesep=.4em,
       xleftmargin=1.2em,xrightmargin=0.2em,
       numbers=left,numbersep=.4em}

\newcommand{\code}[1]{``\texttt{#1}''}
\newcommand{\vb}[1]{\boldsymbol{#1}}
\newcommand{\hm}[1]{#1\nobreak\discretionary{}{\hbox{\ensuremath{#1}}}{}}

\usepackage[pdftex,unicode]{hyperref}
\hypersetup{unicode=true,
pdfauthor={P.V. Moskalev},
pdfsubject={The structure of site percolation models on three-dimensional square lattices},
pdftitle={Moskalev P.V. The structure of site percolation models on 3D square lattices},
pdfkeywords={site percolation, n-dimensional square lattice, non-metric Minkowski distance, R programming language, SPSL package}}

\begin{document}

\begin{titlepage}

\English
\begin{flushleft}\bfseries
UDC 519.676
\end{flushleft}
\begin{flushleft}\bfseries\Large
The structure of site percolation models on three-dimensional square lattices
\end{flushleft}
\begin{flushleft}\bfseries
P.V. Moskalev
\end{flushleft}
\small
Voronezh State Agricultural University (moskalefff@gmail.com)

\paragraph{Abstract.} In this paper we consider the structure of site percolation models on three\-/dimensional square lattices with various shapes of $(1, d)$\-/neighborhood.
For these models, are proposed iso- and anisotropic modifications of the invasion percolation algorithm with $(1, 0)$- and $(1, d)$\-/neighborhoods.
All the above algorithms are special cases of the anisotropic invasion percolation algorithm on the $n$\-/dimensional lattice with a $(1, d)$\-/neighborhood.
This algorithm is the basis for the package SPSL, released under GNU GPL-3 using the free programming language R.

\medskip

\Russian
\begin{flushleft}\bfseries
УДК 519.676
\end{flushleft}
\begin{flushleft}\bfseries\Large
Структура моделей перколяции узлов на трёхмерных квадратных решётках
\end{flushleft}
\begin{flushleft}\bfseries
П.В. Москалев
\end{flushleft}
\small
Воронежский государственный аграрный университет (moskalefff@gmail.com)

\paragraph{Аннотация.}
В работе рассматривается структура моделей перколяции узлов на трёхмерных квадратных решётках при различных формах $(1, \pi)$\-/окрестности.
Для этих моделей предложены изо- и анизотропные модификации алгоритма инвазивной перколяции с $(1, 0)$- и $(1, \pi)$\-/окрестностями.
Все рассмотренные алгоритмы являются частными случаями анизотропного алгоритма инвазивной перколяции на $n$\-/мерной решётке с $(1, \pi)$\-/окрестностью.
Данный алгоритм положен в основу библиотеки SPSL, выпущенной под лицензией GNU GPL-3 с использованием свободного языка программирования R.

\paragraph{Ключевые слова:} перколяция узлов, n-мерная квадратная решётка, неметрическое расстояние Минковского, язык программирования R, библиотека SPSL

\tableofcontents

\normalsize

\end{titlepage}

\section{Введение}

Впервые задачи теории перколяции появились в работах П. Флори и У. Штокмайера \cite{flory.1941.molecular, stockmayer.1943.gel.formation}, посвящённых моделированию полимеризации высокомолекулярных соединений. 
Однако формирование современного математического аппарата и собственной терминологии в исследованиях процессов перколяции принято связывать с публикацией в 1957 году работы С. Бродбента и Дж. Хаммерсли \cite{hammersley.1957.percolation.1}, в которой они рассматривают задачу о протекании некоторой жидкости через случайно-неоднородную проницаемую среду. 

Интересной разновидностью перколяционных моделей является сформулированная Д. Уилкинсоном и Дж. Уиллемсеном задача об инвазивной перколяции \cite{wilkinson.1983.invasion, koplik.1983.displacement}. 
Чаще всего процесс инвазии, или вытеснения, рассматривается как результат взаимодействия двух несмешивающихся жидкостей: первой, уже заполняющей пористую среду, и второй, подаваемой в эту среду под некоторым давлением.
Основное влияние на этот процесс оказывает соотношение градиентов давления в потоках инжектируемой и вытесняемой жидкости, обусловленное действием сил инерции, вязкого трения и межфазного взаимодействия обеих жидкостей со стенками капиллярных каналов.
Существенной особенностью процесса вытеснения является наличие во многих пористых средах тупиковых пор с пониженной связностью, что приводит к захвату во внутрипоровом пространстве некоторого объёма вытесняемой жидкости.

Более простая, но не менее важная разновидность моделей инвазивной перколяции возникает при изучении метода ртутной порометрии.
В экспериментальных исследованиях пористых структур метод инвазивной ртутной порометрии относится к числу наиболее распространённых и хорошо изученных методов, который позволяет оценивать поры с эквивалентным гидравлическим диаметром $d$ от 3,5~нм до 500~мкм.
В его основе лежит высказанная в 1921 году Е. Уошбурном \cite{washburn.1921.porosimetry} идея создания контролируемого перепада давления $\Delta p$ в окружающей пористое тело жидкой ртути для вдавливания некоторого её объёма $\Delta v$ в капилляры последнего.
Квазистатическое увеличение перепада давлений $\Delta p$ позволяет ртути постепенно проникать во всё более мелкие капилляры пористого тела, эквивалентный диаметр $d$ которых будет соответствовать величине вынуждающей силы $\Delta p$, а приращение удельного объёма инжектируемой жидкости $\Delta v$\--- суммарному объёму пор данного диаметра на единицу массы исследуемого образца.
Для снижения влияния на результирующие данные низкой связности тупиковых пор и захвата в поровом пространстве вытесняемых газов и/или жидкостей пористый образец перед испытаниями подвергается вакуумированию, а вдавливаемая жидкость\--- фильтрованию для очистки от посторонних частиц и двойной перегонке для исключения газовыделения в процессе испытаний.
В результате, для адекватного описания процесса ртутной порометрии с помощью моделей инвазивной решёточной перколяции требуется лишь выделение подмножества достижимых узлов решётки, связанного с заданным стартовым подмножеством. 
Отсутствие необходимости в моделировании захвата вытесняемой жидкостью тупиковых и слабосвязных фрагментов порового пространства не только существенно упрощает решаемую задачу, но и позволяет использовать инвазивную решёточную перколяцию без захвата для моделирования внутренней структуры пористой среды.

\section{Общие определения}

Одной из базовых задач, возникающих при моделировании перколяции, является задача выделения подмножества или кластера узлов, непрерывным образом связанных с заданным стартовым подмножеством. 
Простейшая модель инвазивной решёточной перколяции, строится с помощью взвешенного однородного графа или решётки, достижимость произвольного узла которой задается некоторой псевдослучайной величиной $U_i$. 
В том случае, если величины $U_i$ и $U_j$ в соседних узлах решётки при $i=j\pm\varepsilon_k$ взаимно независимы, то говорят о некоррелированной перколяции или перколяции Бернулли.
Величина $\varepsilon_k$ является компонентой сдвигового вектора $\vb\varepsilon(\varepsilon_1, \varepsilon_2, \ldots, \varepsilon_n)$, определяемого формой и радиусом окрестности внутренних узлов решётки.

Из курсов топологии и теории множеств известно \cite{alexandrov.1977.topology}, что ключевое влияние на связность оказывает функция метрики, определяющая расстояния и формирующая $\varepsilon$\-/окрестность некоторой точки $b$. 
Одним из достаточно общих способов определения окрестности произвольной точки $U_{\varepsilon, \pi}(b)$ является использование функции неметрического расстояния Минковского $\rho_\pi(a, b)$:
\begin{gather}\label{eq:mink_metric}
U_{\varepsilon, \pi}(b) = \{a : \rho_\pi(a, b)\leqslant\varepsilon\}, \quad
\rho_\pi(a, b) = \biggl(\sum_{i=1}^n |a_i - b_i|^\pi\biggr)^{1/\pi},
\end{gather}
где $\pi\geqslant 0$\--- показатель неметрического расстояния Минковского (для краткости далее по тексту именуемый просто показателем Минковского); $a(a_1$, $a_2$, \ldots, $a_n)$, $b(b_1$, $b_2$, \ldots, $b_n)$\--- координаты точек $a$ и $b$. 

Применение термина <<неметрическое расстояние>> обусловлено тем, что строгое определение метрики накладывает на функцию \eqref{eq:mink_metric} следующие ограничения:
а)~$\rho_\pi(a, b)=0 \Leftrightarrow a=b$;
б)~$\rho_\pi(a, b)=\rho_\pi(b, a)$;
в)~$\rho_\pi(a, b)\leqslant \rho_\pi(a, c)+\rho_\pi(c, b)$.
Для неметрического расстояния Минковского все три ограничения выполняются лишь при $\pi\geqslant 1$, а на интервале $0\leqslant\pi<1$ знак в третьем неравенстве (неравенстве треугольника) меняется на противоположный $\rho_\pi(a, b)\hm> \rho_\pi(a, c)+\rho_\pi(c, b)$. 
В наших задачах функция неметрического расстояния $\rho_\pi(a, b)$ определяет лишь меру удалённости точек $a$ и $b$ вдоль проходящей через них прямой и используется в \eqref{eq:mink_metric} для определения окрестности $b$ с соответствующим показателем Минковского $\pi$.

В общем случае относительные доли достижимых узлов $p_k$ является компонентами вектора $\vb p(p_1, p_2, \ldots, p_n)$, длина которого соответствует форме и радиусу используемой окрестности внутренних узлов решётки.
При равных компонентах вектора $\vb p$ реализации кластеров будут обладать статистически изотропной структурой, а при неравных\--- структура реализаций станет статистически анизотропной \cite{moskaleff.2007.porous.structures}.

\section[Изотропные кластеры с (1,0)-окрестностью]{Изотропные кластеры с $(1, 0)$\-/окрестностью}

Рассмотрим задачу построения статистически изотропного кластера узлов на трёхмерной квадратной решётке с $(1, 0)$\-/окрестностью фон Неймана.
Среди множества алгоритмов, применяющихся для решения задач перколяции, достаточно высокой эффективностью отличается однопроходный алгоритм, базовые реализации которого в были представлены в работах П. Лиса \cite{leath.1976.cluster.size} и З. Александровиц \cite{alexandrowicz.1980.percolation}. 
Применительно к задаче инвазивной некоррелированной перколяции без захвата основные этапы построения статистически изотропного кластера узлов на трёхмерной решётке с $(1, 0)$\-/окрестностью можно сформулировать следующим образом \cite{moskaleff.jtf.2009.percolation}:
\begin{itemize}
\item[а)]
все узлы перколяционной решётки взвешиваются последовательностью псевдослучайных чисел с равномерным распределением $u_{xyz}\hm\sim \mathbf{U}(0,1)$; недостижимыми считаются те узлы перколяционной решётки, весовой коэффициент $u_{xyz}$ которых больше или равен заданной доли достижимых узлов $p$, а достижимыми\--- узлы, весовой коэффициент которых меньше доли достижимых узлов 
\begin{gather}\label{eq:u<p_ssi30}
u_{xyz}<p;
\end{gather}
\item[б)]
среди узлов решётки формируется некоторое стартовое подмножество и либо все, либо только достижимые узлы стартового подмножества помечаются числовой меткой $l>1$, к примеру $l=2$;
\item[в)]
узлы стартового подмножества объединяются с достижимыми узлами из своего $(1, 0)$\-/периметра, формируемого как объединение $(1, 0)$\-/окрестностей, и помечаются числовой меткой $l$;
\item[г)]
достижимые узлы $(1, 0)$\-/периметра образуют новое стартовое подмножество на следующей итерации;
\item[д)]
пункты (в-г) повторяются до исчерпания достижимых узлов в текущем $(1, 0)$\-/периметре кластера, либо до присоединения к кластеру узлов из заданного целевого подмножества.
\end{itemize}

Описанный алгоритм используется в этом разделе для построения статистически изотропных реализаций кластеров узлов и распределений относительных частот по их выборочной совокупности на трёхмерной квадратной решётке с $(1, 0)$\-/окрестностью фон Неймана.
В листинге~1 показана реализация на языке C общей функции \code{ssTNd()}, обеспечивающей маркировку статистически изо- или анизотропных кластеров узлов, связанных с заданным стартовым подмножеством на $n$\-/мерной квадратной решётке с $(1, \pi)$\-/окрестностью Мура.

\paragraph{Листинг 1.} Реализация функции \code{ssTNd()} на языке C
\begin{Verbatim} 
#include <R.h>
#include <Rinternals.h>
///////////////////////////////////////////////////////////////////
// Функция ssTNd() обеспечивает маркировку статистически изо- или 
// анизотропного кластера узлов, связанного со стартовым подмно-
// жеством на n-мерной квадратной решётке с (1,п)-окрестностью.
///////////////////////////////////////////////////////////////////
// Аргументы:
//   bA - число узлов стартового подмножества;
//   clA - линейные индексы узлов кластера;
//   acA - матрица достижимости узлов решётки;
//   eA, pA - линейные индексы и относительные доли 
//            достижимых узлов для (1,п)-окрестности.
// Переменные:
//   cls, acc, e, b, p - указатели на clA, acA, eA, bA, pA
//   a, *b - индексы узлов из текущего (1,п)-периметра 
//           кластера по вектору cls[]: от, до;
//   c - индекс текущего узла периметра по вектору cls[];
//   h - индекс текущего узла окрестности по вектору e[];
//   n - число узлов, образующих (1,п)-окрестность.
///////////////////////////////////////////////////////////////////
SEXP ssTNd(SEXP pA, SEXP acA, SEXP bA, SEXP eA, SEXP clA) {
  acA = coerceVector(acA, REALSXP);
  clA = coerceVector(clA, INTSXP);
  pA = coerceVector(pA, REALSXP);
  bA = coerceVector(bA, INTSXP);
  eA = coerceVector(eA, INTSXP);
  double *p, *acc;
  int *b, *e, *cls,
      n=length(eA), a=0, db, c, h, ch;
  cls = INTEGER(clA); e = INTEGER(eA); b = INTEGER(bA); 
  acc = REAL(acA); p = REAL(pA); db = *b;
  while (db>0) {              // Пока периметр непуст:
    db = 0;                   // Обнуляем текущий периметр.
    for (c=a; c<*b; c++) {    // Для всех узлов периметра:
      for (h=0; h<n; h++) {   // Для всех узлов окрестности:
        ch = cls[c] + e[h];
        if (acc[ch] < p[h]) { // Если узел достижим
          acc[ch] = 2; db++;  // то помечаем узел и
          cls[*b+db-1] = ch;  // сохраняем его индекс.
        }
      }
    }
    a = *b; *b += db;         // Индексы текущего периметра.
  }
  return(R_NilValue);
}
\end{Verbatim}

В листинге~2 показана реализация на языке R функций \code{ssi30()} и \code{fssi30()}, которые обеспечивают инициализацию переменных, необходимых для корректного вызова функции \code{ssTNd()}.
Функция \code{ssi30()} обеспечивает маркировку статистически изотропного кластера узлов, связанного с заданным стартовым подмножеством \code{set} на трёхмерной квадратной решётке заданного размера \code{x} с $(1, 0)$\-/окрестностью фон Неймана при заданной доле достижимых узлов \code{p}.
Функция \code{fssi30()} обеспечивает расчёт матрицы относительных частот узлов для выборки статистически изотропных реализаций заданного объёма \code{n}, связанных с заданным стартовым подмножеством \code{set} на трёхмерной квадратной решётке заданного размера \code{x} с $(1, 0)$\-/окрестностью фон Неймана при заданной доле достижимых узлов \code{p}.

\paragraph{Листинг 2.} Реализации функций \code{ssi30()} и \code{fssi30()} на языке R
\begin{Verbatim} 
# # # # # # # # # # # # # # # # # # # # # # # # # # # # # # # # # #
# Функции: 
# ssi30()  - обеспечивает маркировку изотропных кластеров узлов на 
#            трёхмерной квадратной решётке с (1,0)-окрестностью;
# fssi30() - обеспечивает построение матрицы относительных частот 
#            для выборки изотропных кластеров узлов на трёхмерной 
#            квадратной решётке с (1,0)-окрестностью.
# # # # # # # # # # # # # # # # # # # # # # # # # # # # # # # # # #
# Аргументы:
# n - объём выборки кластеров;
# x - линейный размер перколяционной решётки; 
# p - относительная доля достижимых узлов решётки;
# set - линейные индексы стартового подмножества;
# all - триггер: "Маркировать все узлы или только достижимые?"
# Переменные:
# e - линейные индексы узлов из (1,0)-окрестности;
# b - длина стартового подмножества узлов.
# Значения:
# acc - матрица достижимости узлов перколяционной решётки;
# rfq - матрица относительных частот узлов перколяционной решётки.
# # # # # # # # # # # # # # # # # # # # # # # # # # # # # # # # # #
ssi30 <- function(x=33, p=0.311608, 
                  set=(x^3+1)/2, all=TRUE) {
  e <- as.integer(c(-1, 1,-x, x,-x^2, x^2))
  p <- as.double(rep(p, length(e)))
  acc <- array(runif(x^3), rep(x, times=3))
  if (!all) set <- set[acc[set] < mean(p)]
  b <- as.integer(length(set))
  cls <- rep(0L, max(p)*x^3 + b*all)
  acc[set] <- 2 
  acc[c(1,x),,] <- acc[,c(1,x),] <- acc[,,c(1,x)] <- 1
  cls[seq_along(set)] <- as.integer(set - 1)
  .Call("ssTNd", p, acc, b, e, cls)   
  return(acc)
}
fssi30 <- function(n=1000, x=33, p=0.311608, 
                   set=(x^3+1)/2, all=TRUE) {
  rfq <- array(0, dim=rep(x, times=3))
  for (i in seq(n))
    rfq <- rfq + (ssi30(x, p, set, all) > 1)
  return(rfq/n)
}
\end{Verbatim}

Приведённые выше реализации были опубликованы автором в составе библиотеки SPSL \cite{moskalev.cran.2012.spsl} под лицензией GNU GPL-3 и с июня 2012 года доступны для свободной загрузки через систему репозиториев CRAN.

\begin{figure}[hbt]\centering
\includegraphics{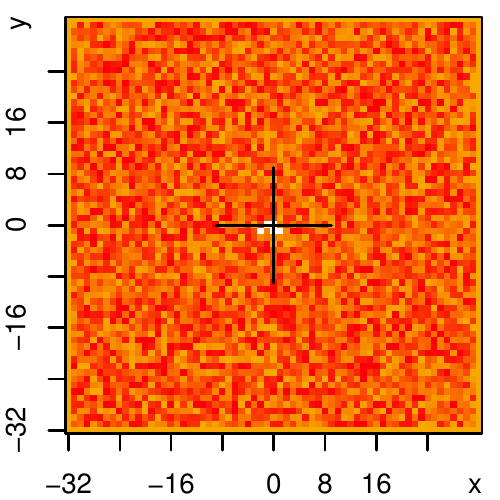}
\includegraphics{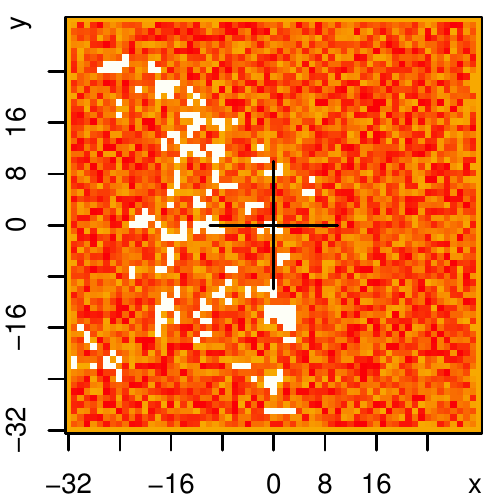}
\includegraphics{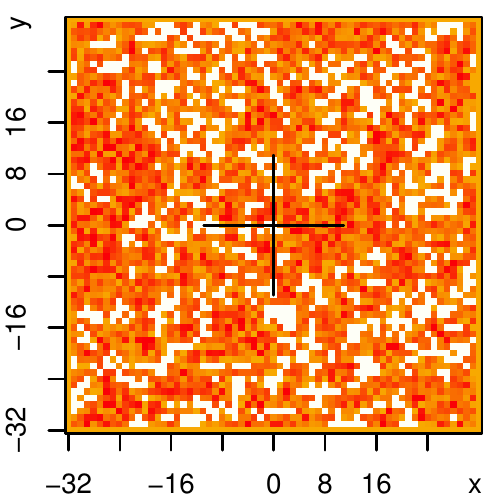}\\[1ex]
\includegraphics{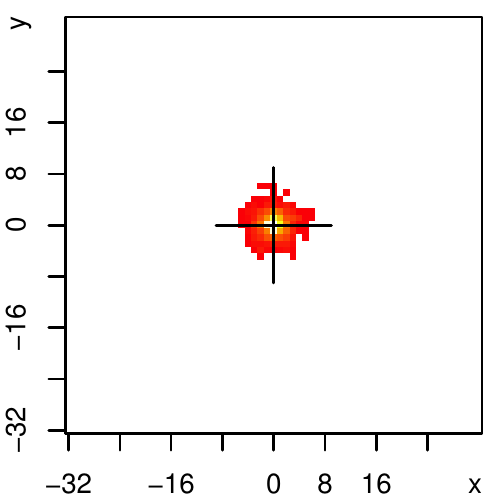}
\includegraphics{fssi30bc}
\includegraphics{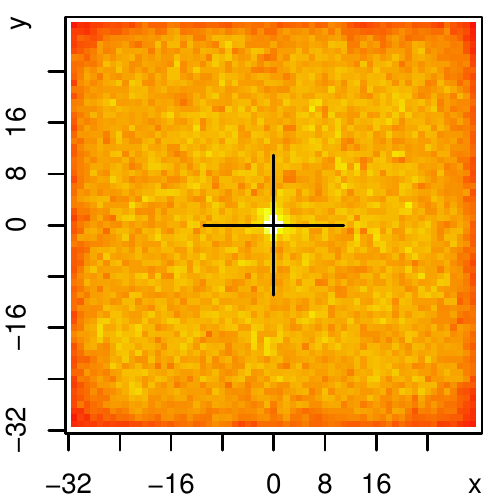}
\caption{\label{pic:ssi30}
Сечения плоскостью $z=0$ изотропных кластеров (верхний ряд) и распределений относительных частот узлов (нижний ряд) с $(1, 0)$\-/окрестностью на трёхмерной квадратной решётке размером $x=65$ узлов для выборки объёмом $m=500$ реализаций при: $p\hm=0,2816<p_c$ (слева); $p=0,3116\approx p_c$ (в центре); $p=0,3416>p_c$ (справа)}
\end{figure}

Примеры построения отдельных реализаций статистически изотропных кластеров и распределений относительных частот на трёхмерной квадратной решётке размером $x=65$ узлов с $(1, 0)$\-/окрестностью фон Неймана и непроницаемыми граничными условиями при различных значениях доли достижимых узлов $p$ показаны на рис.\,\ref{pic:ssi30}.

Реализации и распределения относительных частот, показанные на рис.\,\ref{pic:ssi30} (слева), соответствуют докритическим значениям доли достижимых узлов $p<p_c$.
На рис.\,\ref{pic:ssi30} (в центре) показаны реализации и распределения при околокритических значениях доли достижимых узлов $p\approx p_c$, а на рис.\,\ref{pic:ssi30} (справа)\--- при сверхкритических значениях $p>p_c$.

Для большей наглядности взвешивающая данную решётку последовательность псевдослучайных чисел $u_{xyz}\sim \mathbf{U}(0, 1)$ была зафиксирована. 
Белым цветом в верхнем ряду на рис.\,\ref{pic:ssi30} обозначены узлы, принадлежащие кластеру; оранжевым цветом\--- недостижимые узлы, а красным цветом\--- достижимые узлы, не связанные со стартовым подмножеством.

Красным цветом в нижнем ряду на рис.\,\ref{pic:ssi30} показаны узлы с относительными частотами $w_{xy0}=0,02$, а белым\--- с частотами $w_{xy0}=0,32$; все узлы с частотами, выходящими за пределы данного интервала $w_{xy0}\notin[0,02; 0,32]$, условно не показаны.
Символом ``$+$'' на рис.\,\ref{pic:ssi30} отмечено стартовое подмножество узлов кластера в центре решётки, причём размеры элементов символа относительно центральной точки $(0, 0, 0)$ соответствуют доле достижимых узлов, масштабированной по радиусу перколяционной решётки $\frac{(x-1)p}{2}$.

\section[Изотропные кластеры с (1,п)-окрестностью]{Изотропные кластеры с $(1, \pi)$\-/окрестностью}

В классической задаче о перколяции узлов на трёхмерной квадратной решётке используется единичная окрестность фон Неймана $U_{1, 0}(b)$, соответствующая нулевому показателю Минковского $\pi=0$. 
Для выделенного узла $b$ такая окрестность включает в себя 6 узлов, только одна из координат которых отличается от одноимённой координаты узла $b$ на единицу.
В другом предельном случае при $\pi\to\infty$ единичная окрестность Мура для выделенного узла $b$ на трёхмерной квадратной решётке $U_{1, \infty}(b)$ будет включать в себя 26 узлов, хотя бы одна из координат которых отличается от одноимённой координаты узла $b$ на единицу.

Нетрудно проверить, что единичная окрестность Мура для узла $b$ на трёхмерной квадратной решётке образуется как объединение трёх подмножеств узлов, для которых: а)~только одна; б)~только две; в)~только три из координат отличаются от одноимённой координаты $b$ на единицу.
При промежуточных показателях Минковского $\pi\in(0, \infty)$ мера удалённости $\rho_\pi$ для 6 осевых узлов, образующих окрестность $U_{1, 0}$, будет сохраняться постоянной и равной единице $\rho_{\pi,1}=1$, а для остальных узлов, входящих в окрестность $U_{1, \infty}$, будет убывать от $\rho_{\pi,2}\to\rho_{\pi,3}\to\infty$ при $\pi\to 0+$ до $\rho_{\pi,3}\to\rho_{\pi,2}\to 1+$ при $\pi\to\infty$.
Например, в манхэттенской метрике при $\pi=1$ расстояние до неосевых узлов будет равно $\rho_{1,2}=2$ или $\rho_{1,3}=3$, а в евклидовой метрике при $\pi=2$ расстояние до тех же узлов будет равно $\rho_{2,2}=\sqrt{2}$ или $\rho_{2,3}=\sqrt{3}$. 

Для учёта неметрического расстояния в модели изотропной решёточной перколяции узлов проведём нормировку доли достижимых узлов $p$ в неравенстве \eqref{eq:u<p_ssi30} на меру их удалённости $\rho_\pi$ от текущего узла. 
Тогда итерационный процесс построения реализации перколяционного кластера будет основываться на проверке выполнения весового неравенства 
\begin{gather}\label{eq:u<p/rho_ssi3d}
u_{xyz}<\frac{p}{\rho_\pi}
\end{gather} 
для каждого узла $(x, y, z)$ из единичной окрестности Мура некоторого текущего подмножества узлов, где $u_{xyz}\sim\mathbf{U}(0, 1)$\--- равномерно распределённые на интервале $(0, 1)$ псевдослучайные числа; $p$\--- относительная доля достижимых узлов решётки. 

Также как и в предыдущем случае те узлы, для которых неравенство \eqref{eq:u<p/rho_ssi3d} выполняется, помечаются числовой меткой $l>1$ и образуют текущее подмножество для следующей итерации. 
Условием остановки процесса является появление на очередной итерации пустого текущего подмножества узлов или достижение узла из заданного целевого подмножества.

В листинге~3 показана реализация на языке R функций \code{ssi3d()} и \code{fssi3d()}, которые обеспечивают инициализацию переменных, необходимых для корректного вызова функции \code{ssTNd()}.
Функция \code{ssi3d()} обеспечивает маркировку статистически изотропного кластера узлов, связанного с заданным стартовым подмножеством \code{set} на трёхмерной квадратной решётке заданного размера \code{x} с $(1, \pi)$\-/окрестностью Мура при заданных долях достижимых узлов \code{p0}, \code{p1} и \code{p2}.
Функция \code{fssi3d()} обеспечивает расчёт матрицы относительных частот узлов для выборки статистически изотропных реализаций заданного объёма \code{n}, связанных с заданным стартовым подмножеством \code{set} на трёхмерной квадратной решётке заданного размера \code{x} с $(1, \pi)$\-/окрестностью Мура при заданных долях достижимых узлов \code{p0}, \code{p1} и \code{p2}.
\medskip

\paragraph{Листинг 3.} Реализации функций \code{ssi3d()} и \code{fssi3d()} на языке R
\begin{Verbatim} 
# # # # # # # # # # # # # # # # # # # # # # # # # # # # # # # # # #
# Функции: 
# ssi3d()  - обеспечивает маркировку изотропных кластеров узлов на 
#            трёхмерной квадратной решётке с (1,п)-окрестностью;
# fssi3d() - обеспечивает построение матрицы относительных частот 
#            для выборки изотропных кластеров узлов на трёхмерной 
#            квадратной решётке с (1,п)-окрестностью.
# # # # # # # # # # # # # # # # # # # # # # # # # # # # # # # # # #
# Аргументы:
# n - объём выборки кластеров;
# x - линейный размер перколяционной решётки; 
# p0 - относительная доля достижимых узлов решётки;
# p1 - значение p0, взвешенное на расстояние до узлов из
#      двумерной (1,п)-окрестности;
# p2 - значение p0, взвешенное на расстояние до узлов из
#      трёхмерной (1,п)-окрестности;
# set - линейные индексы стартового подмножества;
# all - триггер: "Маркировать все узлы или только достижимые?"
# Переменные:
# e0 - линейные индексы узлов из (1,0)-окрестности;
# e1 - линейные индексы узлов из двумерной (1,п)-окрестности;
# e2 - линейные индексы узлов из трёхмерной (1,п)-окрестности;
# b - длина стартового подмножества узлов.
# Значения:
# acc - матрица достижимости узлов перколяционной решётки;
# rfq - матрица относительных частот узлов перколяционной решётки.
# # # # # # # # # # # # # # # # # # # # # # # # # # # # # # # # # #
ssi3d <- function(x=33, p0=0.2, p1=p0/2, p2=p0/3,
                  set=(x^3+1)/2, all=TRUE) {
  e0 <- c(-1, 1,-x, x,-x^2, x^2)
  e1 <- colSums(matrix(e0[c(
    1,3, 2,3, 1,4, 2,4,
    1,5, 2,5, 1,6, 2,6,
    3,5, 4,5, 3,6, 4,6)], nrow=2))
  e2 <- colSums(matrix(e0[c(
    1,3,5, 2,3,5, 1,4,5, 2,4,5,
    1,3,6, 2,3,6, 1,4,6, 2,4,6)], nrow=3))
  e <- as.integer(c(e0,e1,e2))
  p0 <- rep(p0, length(e0))
  p1 <- rep(p1, length(e1))
  p2 <- rep(p2, length(e2))
  p <- as.double(c(p0,p1,p2))  
  acc <- array(runif(x^3), rep(x, times=3))
  if (!all) set <- set[acc[set] < mean(p)]
  b <- as.integer(length(set))
  cls <- rep(0L, max(p)*x^3 + b*all)
  acc[set] <- 2 
  acc[c(1,x),,] <- acc[,c(1,x),] <- acc[,,c(1,x)] <- 1
  cls[seq_along(set)] <- as.integer(set - 1)
  .Call("ssTNd", p, acc, b, e, cls)  
  return(acc)
}
fssi3d <- function(n=1000, x=33, p0=0.2, p1=p0/2, p2=p0/3, 
                   set=(x^3+1)/2, all=TRUE) {
  rfq <- array(0, dim=rep(x, times=3))
  for (i in seq(n))
    rfq <- rfq + (ssi3d(x, p0, p1, p2, set, all) > 1)
  return(rfq/n)
}
\end{Verbatim}

Приведённые выше реализации функций \code{ssi3d()} и \code{fssi3d()} также вошли в состав библиотеки SPSL \cite{moskalev.cran.2012.spsl}, опубликованной автором под лицензией GNU GPL-3.

\begin{figure}[hbt]\centering
\includegraphics{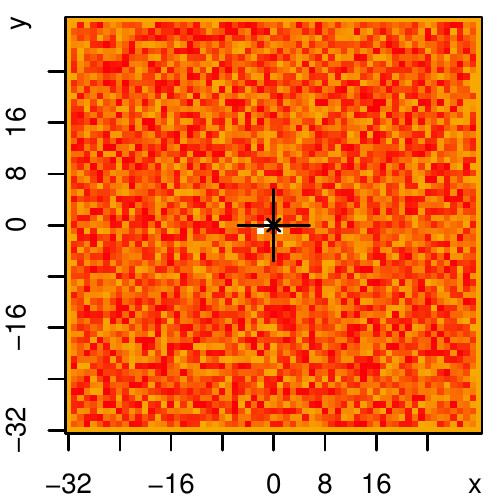}
\includegraphics{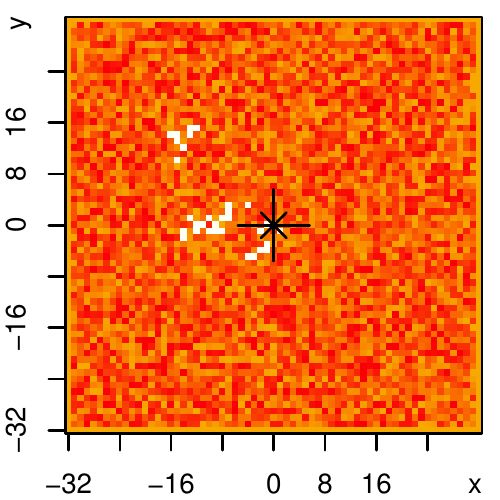}
\includegraphics{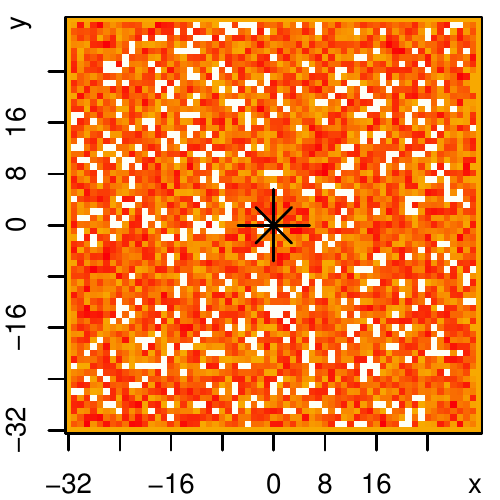}\\[1ex]
\includegraphics{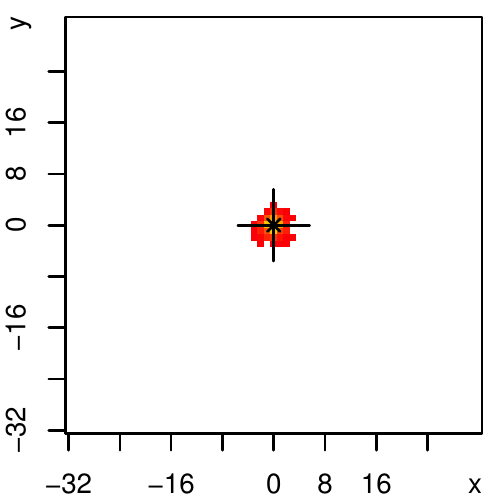}
\includegraphics{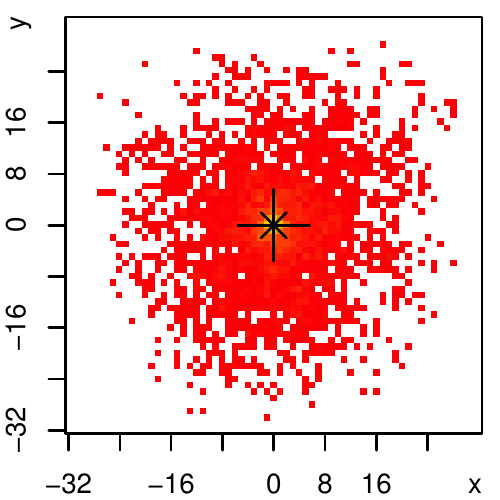}
\includegraphics{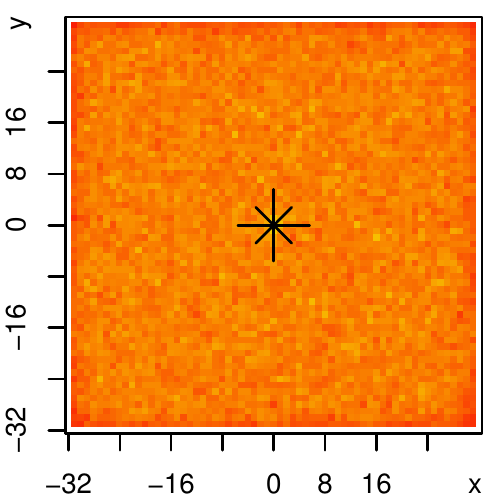}
\caption{\label{pic:ssi3d}
Сечения плоскостью $z=0$ изотропных кластеров (верхний ряд) и распределений относительных частот узлов (нижний ряд) с $(1, \pi)$\-/окрестностью на трёхмерной квадратной решётке размером $x=65$ узлов для выборки объёмом $m=500$ реализаций при $p=0,175$ и: $\pi=0,5$ (слева); $\pi=1$ (в центре); $\pi=2$ (справа)}
\end{figure}

Примеры построения отдельных реализаций статистически изотропных кластеров и распределений относительных частот на трёхмерной квадратной решётке размером $x=65$ узлов с $(1, \pi)$\-/окрестностью Мура и непроницаемыми граничными условиями при фиксированной доле достижимых узлов $p=0,175$ и различных значениях показателя Минковского $\pi$ приведены на рис.\,\ref{pic:ssi3d}.

На рис.\,\ref{pic:ssi3d} (слева) показаны реализации и распределения относительных частот при значениях показателя Минковского $\pi=0,5$, соответствующих неметрическим расстояниям.
На рис.\,\ref{pic:ssi3d} (в центре) показаны реализации и распределения относительных частот при значениях $\pi=1$, соответствующих манхеттенской метрике, а на рис.\,\ref{pic:ssi3d} (справа)\--- при значениях $\pi=2$, соответствующих евклидовой метрике.

Остальные параметры для отображения отдельных реализаций в верхнем ряду на рис.\,\ref{pic:ssi3d} и распределений относительных частот в нижнем ряду на рис.\,\ref{pic:ssi3d} выбраны идентичными примерам из предыдущего раздела.
Символом ``$\lefteqn{\times}+$'' на рис.\,\ref{pic:ssi3d} отмечено стартовое подмножество узлов кластера в центре решётки, причём размеры элементов символов относительно центральной точки $(0, 0, 0)$ соответствуют долям достижимых узлов и их попарным средним, нормированным на меру удалённости узла в $(1, \pi)$\-/окрестности Мура и масштабированным по радиусу перколяционной решётки $\frac{(x-1)p}{2\rho_\pi}$.

\section[Анизотропные кластеры с (1,0)-окрестностью]{Анизотропные кластеры с $(1, 0)$\-/окрестностью}

Рассмотрим задачу о построении статистически анизотропного кластера узлов на трёхмерной квадратной решётке с $(1, 0)$\-/окрестностью фон Неймана.
С алгоритмической точки зрения это означает векторизацию изотропного весового неравенства \eqref{eq:u<p_ssi30} по узлам, образующим $(1, 0)$\-/окрестность фон Неймана \cite{moskaleff.mce.2013.ssTNd.perc}
\begin{equation}\label{eq:u<p_ssa30}
u_{xyz}<p_k
\quad\text{для}\quad
k=1, 2,\ldots, n,
\end{equation}
где $p_k$\--- компоненты вектора долей достижимых узлов $\vb p$ размерности $n=6$ для трёхмерной решётки.
При $p_1=p_2=\ldots=p_n$ кластер узлов будет статистически изотропным, в противном случае\--- неравные компоненты вектора $\vb p$ приведут к появлению ненулевой меры статистической анизотропии.

Для количественной оценки статистической анизотропии кластера воспользуемся евклидовой нормой разности векторов $L_2=||\vb p-\langle\vb p\rangle||$, где $\langle\vb p\rangle$\--- усреднённый по компонентам вектор долей достижимых узлов $\vb p$.
Тогда в изотропном случае при $\vb p=\langle\vb p\rangle$ мера анизотропии будет равна нулю $L_2=0$, а в анизотропном случае при $\vb p\neq\langle\vb p\rangle$\--- строго больше нуля $L_2>0$.

Также как и в изотропном случае те узлы, для которых неравенство \eqref{eq:u<p_ssa30} выполняется, помечаются числовой меткой $l>1$ и образуют текущее подмножество для следующей итерации. 
Условием остановки процесса является появление на очередной итерации пустого текущего подмножества узлов или достижение узла из заданного целевого подмножества.

В листинге~4 показана реализация на языке R функций \code{ssa30()} и \code{fssa30()}, которые обеспечивают инициализацию переменных, необходимых для корректного вызова функции \code{ssTNd()}.
Функция \code{ssa30()} обеспечивает маркировку статистически анизотропного кластера узлов, связанного с заданным стартовым подмножеством \code{set} на трёхмерной квадратной решётке заданного размера \code{x} с $(1, 0)$\-/окрестностью фон Неймана при заданных компонентах вектора долей достижимых узлов \code{p}.
Функция \code{fssa30()} обеспечивает расчёт матрицы относительных частот узлов для выборки статистически анизотропных реализаций заданного объёма \code{n}, связанных с заданным стартовым подмножеством \code{set} на трёхмерной квадратной решётке заданного размера \code{x} с $(1, 0)$\-/окрестностью фон Неймана при заданных компонентах вектора долей достижимых узлов \code{p}.
\medskip

\paragraph{Листинг 4.} Реализации функций \code{ssa30()} и \code{fssa30()} на языке R
\begin{Verbatim} 
# # # # # # # # # # # # # # # # # # # # # # # # # # # # # # # # # #
# Функции: 
# ssa30()  - обеспечивает маркировку анизотропных кластеров узлов на 
#            трёхмерной квадратной решётке с (1,0)-окрестностью;
# fssa30() - обеспечивает построение матрицы относительных частот 
#            для выборки анизотропных кластеров узлов на трёхмерной 
#            квадратной решётке с (1,0)-окрестностью.
# # # # # # # # # # # # # # # # # # # # # # # # # # # # # # # # # #
# Аргументы:
# n - объём выборки кластеров;
# x - линейный размер перколяционной решётки; 
# p - вектор относительных долей достижимых узлов по основным 
#     направлениям решётки: -x, +x, -y, +y, -z, +z;
# set - линейные индексы стартового подмножества;
# all - триггер: "Маркировать все узлы или только достижимые?"
# Переменные:
# e - линейные индексы узлов из (1,0)-окрестности;
# b - длина стартового подмножества узлов.
# Значения:
# acc - матрица достижимости узлов перколяционной решётки;
# rfq - матрица относительных частот узлов перколяционной решётки.
# # # # # # # # # # # # # # # # # # # # # # # # # # # # # # # # # #
ssa30 <- function(x=33, p=runif(6, max=0.6), 
                  set=(x^3+1)/2, all=TRUE) {
  e <- as.integer(c(-1, 1,-x, x,-x^2, x^2))
  p <- as.double(p)
  acc <- array(runif(x^3), rep(x,3))
  if (!all) set <- set[acc[set] < mean(p)]
  b <- as.integer(length(set))
  cls <- rep(0L, max(p)*x^3 + b*all)
  acc[set] <- 2 
  acc[c(1,x),,] <- acc[,c(1,x),] <- acc[,,c(1,x)] <- 1
  cls[seq_along(set)] <- as.integer(set - 1)
  .Call("ssTNd", p, acc, b, e, cls) 
  return(acc)
}
fssa30 <- function(n=1000, x=33, p=runif(6, max=0.6), 
                   set=(x^3+1)/2, all=TRUE) {
  rfq <- array(0, dim=rep(x, times=3))
  for (i in seq(n))
    rfq <- rfq + (ssa30(x, p, set, all) > 1)
  return(rfq/n)
}
\end{Verbatim}

Приведённые выше реализации функций \code{ssa30()} и \code{fssa30()} вошли в состав библиотеки SPSL \cite{moskalev.cran.2012.spsl}, опубликованной автором под лицензией GNU GPL-3.

\begin{figure}[!ht]\centering
\includegraphics{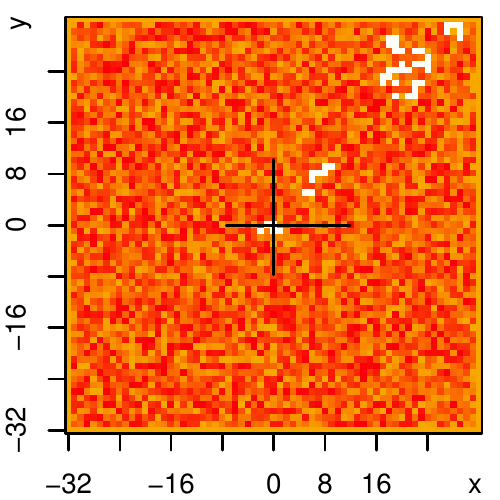}
\includegraphics{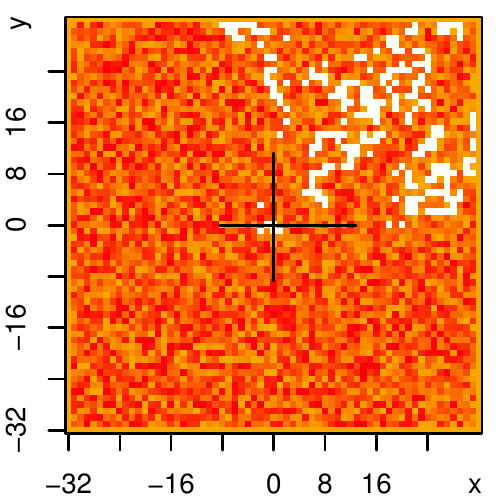}
\includegraphics{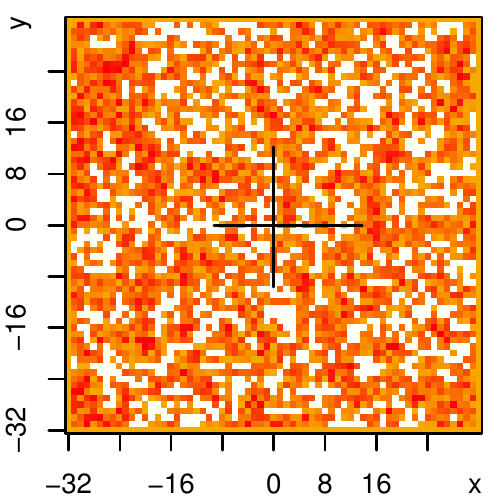}\\[1ex]
\includegraphics{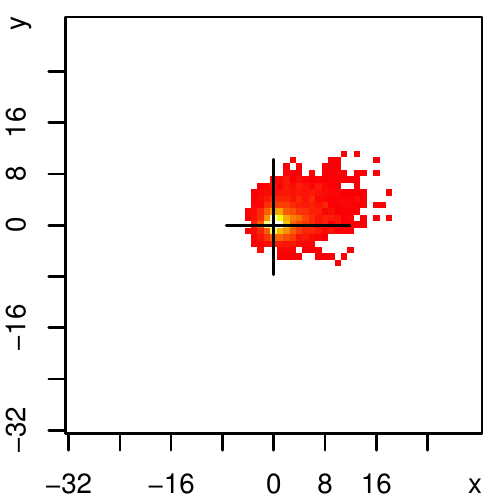}
\includegraphics{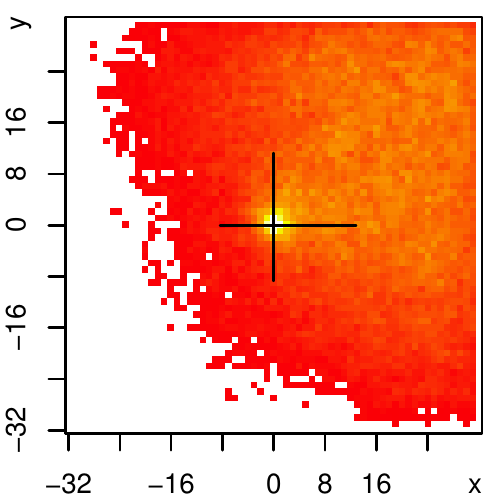}
\includegraphics{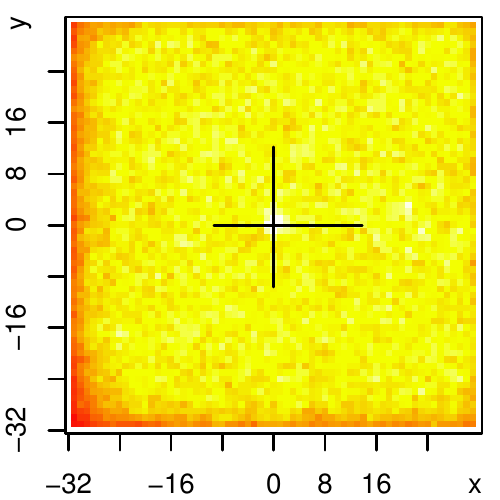}
\caption{\label{pic:ssa30}
Сечения плоскостью $z=0$ анизотропных кластеров (верхний ряд) и распределений относительных частот узлов (нижний ряд) с $(1, 0)$\-/окрестностью фон Неймана на трёхмерной квадратной решётке размером $x=65$ узлов для выборки объёмом $m=500$ реализаций при:
$\vb p_{1}=(0,232$; $0,372$; $0,242$; $0,322$; $0,282$; $0,282)$ (слева);
$\vb p_{2}=(0,262$; $0,402$; $0,272$; $0,352$; $0,312$; $0,312)$ (в центре);
$\vb p_{3}=(0,292$; $0,432$; $0,302$; $0,382$; $0,342$; $0,342)$ (справа)}
\end{figure}

Примеры построения отдельных реализаций статистически анизотропных кластеров и распределений относительных частот на трёхмерной квадратной решётке размером $x=65$ узлов с $(1, 0)$\-/окрестностью фон Неймана и непроницаемыми граничными условиями при различных векторах долей достижимых узлов $\vb p$ показаны на рис.\,\ref{pic:ssa30}.

Основные параметры для отображения отдельных реализаций в верхнем ряду на рис.\,\ref{pic:ssa30} и распределений относительных частот в нижнем ряду на рис.\,\ref{pic:ssa30} выбраны идентичными примерам из предыдущих разделов.
Символом ``$+$'' на рис.\,\ref{pic:ssa30} отмечено стартовое подмножество узлов кластера в центре решётки, причём размеры элементов символов относительно центральной точки $(0, 0, 0)$ соответствуют компонентам вектора долей достижимых узлов, масштабированным по радиусу перколяционной решётки $\frac{(x-1)\vb p}{2}$.

\section[Анизотропные кластеры с (1,п)-окрестностью]{Анизотропные кластеры с $(1, \pi)$\-/окрестностью}

Рассмотрим задачу о построении статистически анизотропного кластера узлов на трёхмерной квадратной решётке с $(1, \pi)$\-/окрестностью Мура.
С алгоритмической точки зрения это означает нормирование векторизованного весового неравенства \eqref{eq:u<p_ssa30} на неметрическое расстояние Минковского $\rho_\pi$ до данного узла с текущим показателем $\pi$ \cite{moskaleff.mce.2013.ssTNd.perc}: 
\begin{equation}\label{eq:u<p/rho_ssa3d}
u_{xyz}<\frac{p_k}{\rho_\pi}
\quad\text{для}\quad 
k=1, 2,\ldots, n,
\end{equation}
где $p_k$\--- компоненты вектора долей достижимых узлов $\vb p$ размерности $n=26$ для трёхмерной решётки.
$(1, \pi)$\-/окрестность Мура на трёхмерной квадратной решётке образуется как объединение трёх подмножеств узлов, для которых: а)~только одна; б)~только две; в)~только три из координат отличаются от одноимённой координаты выделенного узла на единицу.
Нормирующий делитель взвешивающего неравенства $\rho_\pi$ для элементов этих подмножеств будет принимать одно из трёх значений: а)~$\rho_\pi=1^{1/\pi}=1$; б)~$\rho_\pi=2^{1/\pi}=\sqrt[\pi]{2}$; в)~$\rho_\pi=3^{1/\pi}=\sqrt[\pi]{3}$.

Также как и в предыдущих случаях те узлы, для которых указанное неравенство выполняется, помечаются числовой меткой $l>1$ и образуют текущее подмножество для следующей итерации. 
Условием остановки процесса является появление на очередной итерации пустого текущего подмножества узлов или достижение узла из заданного целевого подмножества.

В листинге~5 показана реализация на языке R функций \code{ssa3d()} и \code{fssa3d()}, которые обеспечивают инициализацию переменных, необходимых для корректного вызова функции \code{ssTNd()}.
Функция \code{ssa3d()} обеспечивает маркировку статистически анизотропного кластера узлов, связанного с заданным стартовым подмножеством \code{set} на трёхмерной квадратной решётке заданного размера \code{x} с $(1, \pi)$\-/окрестностью Мура при заданных компонентах векторов долей достижимых узлов \code{p0}, \code{p1} и \code{p2}.
Функция \code{fssa3d()} обеспечивает расчёт матрицы относительных частот узлов для выборки статистически анизотропных реализаций заданного объёма \code{n}, связанных с заданным стартовым подмножеством \code{set} на трёхмерной квадратной решётке заданного размера \code{x} с $(1, \pi)$\-/окрестностью Мура при заданных компонентах векторов долей достижимых узлов \code{p0}, \code{p1} и \code{p2}.
\clearpage

\paragraph{Листинг 5.} Реализации функций \code{ssa3d()} и \code{fssa3d()} на языке R
\begin{Verbatim} 
# # # # # # # # # # # # # # # # # # # # # # # # # # # # # # # # # #
# Функции: 
# ssa3d()  - обеспечивает маркировку анизотропных кластеров узлов на 
#            трёхмерной квадратной решётке с (1,п)-окрестностью;
# fssa3d() - обеспечивает построение матрицы относительных частот 
#            для выборки анизотропных кластеров узлов на трёхмерной 
#            квадратной решётке с (1,п)-окрестностью.
# # # # # # # # # # # # # # # # # # # # # # # # # # # # # # # # # #
# Аргументы:
# n - объём выборки кластеров;
# x - линейный размер перколяционной решётки; 
# p0 - вектор относительных долей достижимых узлов по основным 
#      направлениям решётки: -x, +x, -y, +y, -z, +z;
# p1 - парные усреднённые комбинации компонент p0, взвешенные на 
#      расстояние до узлов из двумерной (1,п)-окрестности;
# p2 - тройные усреднённые комбинации компонент p0, взвешенные на 
#      расстояние до узлов из трёхмерной (1,п)-окрестности;
# set - линейные индексы стартового подмножества;
# all - триггер: "Маркировать все узлы или только достижимые?"
# Переменные:
# e0 - линейные индексы узлов из (1,0)-окрестности;
# e1 - линейные индексы узлов из двумерной (1,п)-окрестности;
# e2 - линейные индексы узлов из трёхмерной (1,п)-окрестности;
# b - длина стартового подмножества узлов.
# Значения:
# acc - матрица достижимости узлов перколяционной решётки;
# rfq - матрица относительных частот узлов перколяционной решётки.
# # # # # # # # # # # # # # # # # # # # # # # # # # # # # # # # # #
ssa3d <- function(x=33, 
                  p0=runif(6, max=0.4),
                  p1=colMeans(matrix(p0[c(
                    1,3, 2,3, 1,4, 2,4,
                    1,5, 2,5, 1,6, 2,6,
                    3,5, 4,5, 3,6, 4,6)], nrow=2))/2,
                  p2=colMeans(matrix(p0[c(
                    1,3,5, 2,3,5, 1,4,5, 2,4,5,
                    1,3,6, 2,3,6, 1,4,6, 2,4,6)], nrow=3))/3,                  
                  set=(x^3+1)/2, all=TRUE) {
  e0 <- c(-1, 1,-x, x,-x^2, x^2)
  e1 <- colSums(matrix(e0[c(
    1,3, 2,3, 1,4, 2,4, 
    1,5, 2,5, 1,6, 2,6,
    3,5, 4,5, 3,6, 4,6)], nrow=2))
  e2 <- colSums(matrix(e0[c(
    1,3,5, 2,3,5, 1,4,5, 2,4,5,
    1,3,6, 2,3,6, 1,4,6, 2,4,6)], nrow=3))
  e <- as.integer(c(e0,e1,e2))
  p <- as.double(c(p0,p1,p2))
  acc <- array(runif(x^3), rep(x, times=3))
  if (!all) set <- set[acc[set] < mean(p)]
  b <- as.integer(length(set))
  cls <- rep(0L, max(p)*x^3 + b*all)
  acc[set] <- 2 
  acc[c(1,x),,] <- acc[,c(1,x),] <- acc[,,c(1,x)] <- 1
  cls[seq_along(set)] <- as.integer(set - 1)
  .Call("ssTNd", p, acc, b, e, cls) 
  return(acc)
}
fssa3d <- function(n=1000, x=33, 
                   p0=runif(6, max=0.4),
                   p1=colMeans(matrix(p0[c(
                     1,3, 2,3, 1,4, 2,4,
                     1,5, 2,5, 1,6, 2,6,
                     3,5, 4,5, 3,6, 4,6)], nrow=2))/2,
                   p2=colMeans(matrix(p0[c(
                     1,3,5, 2,3,5, 1,4,5, 2,4,5,
                     1,3,6, 2,3,6, 1,4,6, 2,4,6)], nrow=3))/3,
                   set=(x^3+1)/2, all=TRUE) {
  rfq <- array(0, dim=rep(x, times=3))
  for (i in seq(n))
    rfq <- rfq + (ssa3d(x, p0, p1, p2, set, all) > 1)
  return(rfq/n)
}
\end{Verbatim}

Приведённые выше реализации функций \code{ssa3d()} и \code{fssa3d()} вошли в состав библиотеки SPSL \cite{moskalev.cran.2012.spsl}, опубликованной автором под лицензией GNU GPL-3.

\begin{figure}[!ht]\centering
\includegraphics{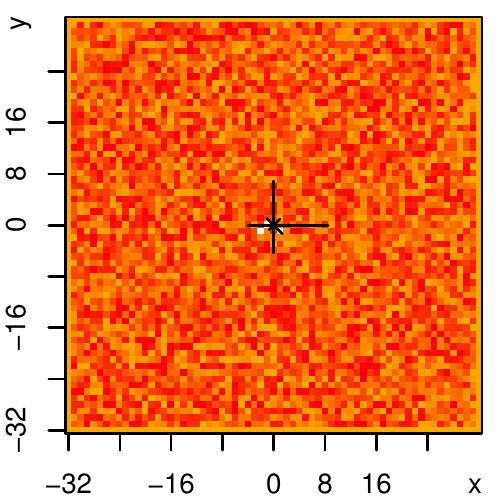}
\includegraphics{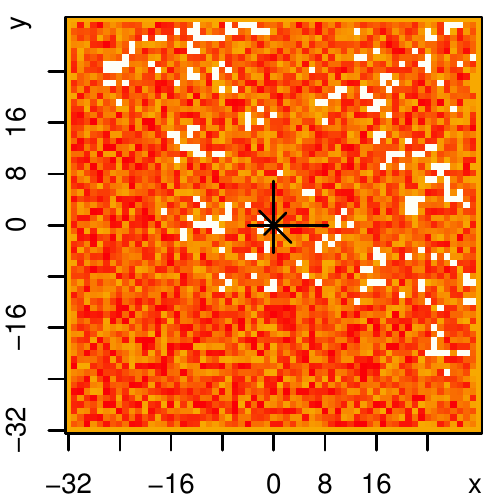}
\includegraphics{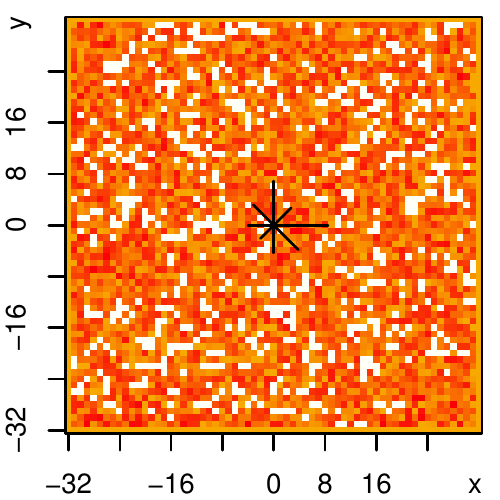}\\[1ex]
\includegraphics{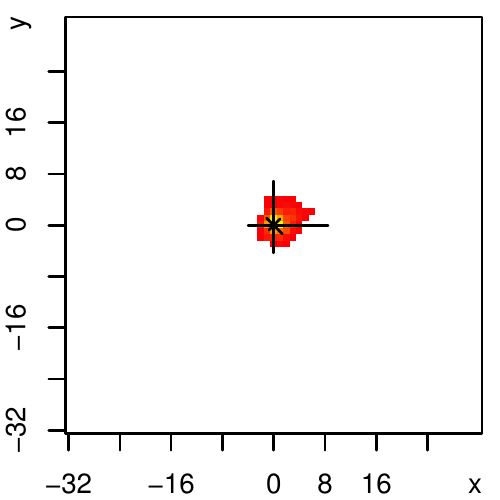}
\includegraphics{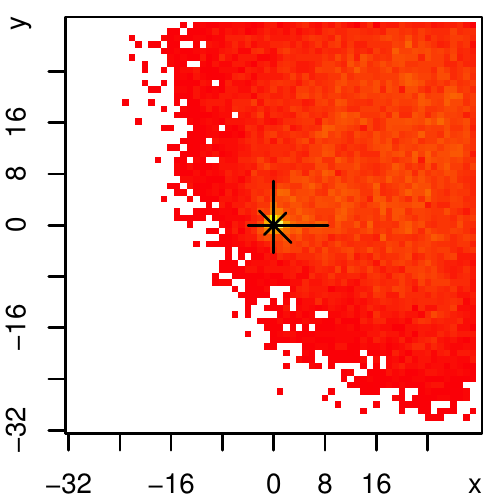}
\includegraphics{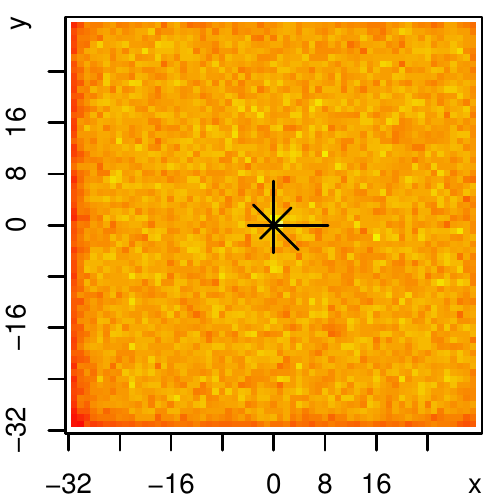}
\caption{\label{pic:ssa3d}
Сечения плоскостью $z=0$ анизотропных кластеров (верхний ряд) и распределений относительных частот узлов (нижний ряд) с $(1, \pi)$\-/окрестностью на трёхмерной квадратной решётке размером $x=65$ узлов для выборки объёмом $m=500$ реализаций при $\vb p=(0,125$; $0,265$; $0,135$; $0,215$; $0,175$; $0,175)$ и: $\pi=0,5$ (слева); $\pi=1$ (в центре); $\pi=2$ (справа)}
\end{figure}

Примеры построения отдельных реализаций статистически анизотропных кластеров и распределений относительных частот на трёхмерной квадратной решётке размером $x=65$ узлов с $(1, \pi)$\-/окрестностью Мура и непроницаемыми граничными условиями при фиксированном векторе долей достижимых узлов $\vb p$ по основным направлениям перколяционной решётки: $\pm\vb i$, $\pm\vb j$, $\pm\vb k$ и различных значениях показателя Минковского $\pi$ приведены на рис.\,\ref{pic:ssa3d}.

На рис.\,\ref{pic:ssa3d} (слева) показаны реализации и распределения относительных частот при значениях показателя Минковского $\pi=0,5$, соответствующих неметрическим расстояниям.
На рис.\,\ref{pic:ssa3d} (в центре) показаны реализации и распределения относительных частот при значениях $\pi=1$, соответствующих манхеттенской метрике, а на рис.\,\ref{pic:ssa3d} (справа)\--- при значениях $\pi=2$, соответствующих евклидовой метрике.

Основные параметры для отображения отдельных реализаций в верхнем ряду на рис.\,\ref{pic:ssa30} и распределений относительных частот в нижнем ряду на рис.\,\ref{pic:ssa30} выбраны идентичными примерам из предыдущих разделов.
Символом ``$\lefteqn{\times}+$'' на рис.\,\ref{pic:ssa3d} отмечено стартовое подмножество узлов кластера в центре решётки, причём размеры элементов символов относительно центральной точки $(0, 0, 0)$ соответствуют компонентам вектора долей достижимых узлов и их попарным средним, нормированным на меру их удалённости в $(1, \pi)$\-/окрестности Мура и масштабированным по радиусу перколяционной решётки $\frac{(x-1)\vb p}{2\rho_\pi}$.

\section{Заключение}

Сопоставление структуры рассмотренных выше моделей перколяции узлов на трёхмерной квадратной решётке показывает, что изотропная модель на решётке с $(1, 0)$\-/окрестностью фон Неймана является частным случаем как изотропной модели с $(1, \pi)$\-/окрестностью Мура, так и анизотропной модели с $(1, 0)$\-/окрестностью фон Неймана.
В свою очередь, изо- и анизотропная модели с $(1, 0)$\-/окрестностью фон Неймана, а также изотропная модель с $(1, \pi)$\-/окрестностью Мура являются частными случаями анизотропной модели с $(1, \pi)$\-/окрестностью Мура, причём размерность решётки на указанные отношения существенной роли не влияет.
Это приводит к выводу о существовании иерархической структуры моделей перколяции узлов на $n$\-/мерных квадратных решётках, допускающих описание в рамках универсального алгоритма, реализация которого приведена в листинге~1.

Сравнение статистически изотропных распределений относительных частот для инвариантно взвешенных решёток с единичными окрестностями фон Неймана и Мура в нижнем ряду на рис.\,\ref{pic:ssi30} и \ref{pic:ssi3d} показывает, что вероятность появления перколяционного кластера $P_\infty(p, \pi)$ является возрастающей функцией как по доле достижимых узлов $p$, так и по показателю Минковского $\pi$.
Следовательно, при возрастании значений показателя $\pi$ на решётке с окрестностью Мура критическое значение доли достижимых узлов $p_c(\pi)$ будет снижаться.

Действительно, сравнивая показанные в нижнем ряду слева на рис.\,\ref{pic:ssi30} и \ref{pic:ssi3d} сечения плоскостью $z=0$ статистически изотропных распределений относительных частот узлов с единичными окрестностями фон Неймана и Мура нетрудно заметить, что последний случай соответствует докритическому значению для трёхмерной решётки $p=0,175<p_c(\pi=0,5)$.
Тогда показанные в центре и справа в нижнем ряду на рис.\,\ref{pic:ssi3d} распределения соответствуют около- и сверхкритическим значениям долей достижимых узлов для трёхмерной решётки: $p=0,175\approx p_c(\pi=1)$ и $p\hm=0,175<p_c(\pi=2)$.

Используя сечения плоскостью $z=0$ статистически анизотропных распределений относительных частот, представленных в нижнем ряду на рис.\,\ref{pic:ssa30}, нетрудно заметить, что основное влияние на протекание в заданном направлении перколяционной решётки $\pm\vb i$, $\pm\vb j$, $\pm\vb k$ оказывают значения соответствующих компонент векторов $\vb p_3$.
К примеру, кластеры распределении относительных частот, показанном в нижнем ряду слева на рис.\,\ref{pic:ssa30} обладают структурой, сверхкритической в направлениях орт $\vb i$, $\vb j$ и докритической в направлениях $-\vb i$, $-\vb j$.
Действительно, сравнивая компоненты вектора $\vb p_1=(0,232$; $0,372$; $0,242$; $0,322$; $0,282$; $0,282)$ с порогом перколяции узлов на трёхмерной квадратной решётке $p_c\approx 0,311$ можно записать, что доли достижимых узлов имеют докритические значения $p_{11}<p_{13}<p_c$ в направлениях орт $\vb i$, $\vb j$, и сверхкритические $p_{12}>p_{14}>p_c$ в направлениях $-\vb i$, $-\vb j$.
При этом протекания ни в одном из этих направлений не наблюдается, поскольку доли достижимых узлов в ортогональных направлениях $\pm\vb k$ также имеют докритические значения $p_{15}=p_{16}<p_c$.

\end{document}